\documentclass[twocolumn, showpacs,amsmath,amssymb]{revtex4}
\usepackage{graphicx}
\usepackage{dcolumn}
\usepackage{bm}
\usepackage{stackrel}
\usepackage{amsmath}
\usepackage{amsfonts}
\usepackage{amssymb}
\usepackage{amsthm}
\usepackage{epstopdf}
\usepackage{longtable}
\usepackage{array}
\usepackage{pifont}
\usepackage{latexsym}
\usepackage[latin1]{inputenc}
\usepackage{bm}
\usepackage{textcomp}
\usepackage{subfigure}
\usepackage{braket}
\usepackage{wasysym}
\usepackage{ulem}
\usepackage{color, soul}
\newcolumntype{.}{D{.}{.}{-1}}

\begin{document}

\author{T. Zanon-Willette$^{1}$\footnote{E-mail address: thomas.zanon@upmc.fr\\}, E. de Clercq$^{2}$, E. Arimondo$^{3}$}
\affiliation{$^{1}$LERMA, Observatoire de Paris, PSL Research University, CNRS, Sorbonne Universités, UPMC Univ. Paris 06, F-75005, Paris, France}
\affiliation{$^{2}$LNE-SYRTE, Observatoire de Paris, PSL Research University, CNRS, Sorbonne Universités, UPMC Univ. Paris 06, 61 avenue de l'Observatoire, 75014 Paris, France}
\affiliation{$^{3}$Dipartimento di Fisica "E. Fermi", Universit\`a di Pisa, Lgo. B. Pontecorvo 3, 56122 Pisa, Italy}

\date{\today}

\preprint{APS/123-QED}

\title{Probe light-shift elimination in Generalized Hyper-Ramsey quantum clocks}

\begin{abstract}
We present a new interrogation scheme for the next generation of  quantum clocks to suppress frequency-shifts induced by laser probing fields themselves based on Generalized Hyper-Ramsey resonances.
Sequences of composite laser pulses with specific selection of phases, frequency detunings and durations are combined to generate a very efficient and robust frequency locking signal with almost a perfect elimination of the light-shift from off resonant states and to decouple the unperturbed frequency measurement from the laser's intensity. The frequency lock point generated from synthesized error signals using either $\pi/4$ or $3\pi/4$ laser phase-steps during the intermediate pulse is tightly protected against large laser pulse area variations and errors in potentially applied frequency shift compensations.
Quantum clocks based on weakly allowed or completely forbidden optical transitions in atoms, ions, molecules and nuclei will benefit from these hyper-stable laser frequency stabilization schemes to reach relative accuracies below the 10$^{-18}$ level.
\end{abstract}


\maketitle

\indent The next generation of optical-frequency standards based on an ensemble of neutral atoms or a single-ion will provide new very stringent tests in metrology, applied and fundamental physics \cite{Ludlow:2015}.
Atomic clocks using cold fermionic species trapped in 1D optical lattices \cite{Derevianko:2011}, now surpassing the accuracy of caesium atomic fountains, have clearly demonstrated the potential to establish a breakthrough in ultra-high precision measurement achieving soon a relative 10$^{-18}$ level of accuracy. For both bosonic and fermionic species, the cold collisions and the light-shifts contributions from the optical lattice or from the probe laser have been intensively explored and mitigated \cite{Akatsuka:2010,Ludlow:2011}. In a recent investigation of the fermionic $^{87}$Sr optical lattice clock \cite{Nicholson:2015}, the lattice and the probe laser light-shift corrections were characterized at the $1\times10^{-18}$ relative level of uncertainty.
Instead probe light-shifts represent a non negligeable issue for clocks based on bosonic neutral atoms with forbidden dipole transitions activated by mixing static magnetic field with a single laser \cite{Taichenachev:2006,Barber:2006,Baillard:2007,Kulosa:2015}, for clocks with magic-wave induced transition in even isotopes \cite{Ovsiannikov:2007}, with E1-M1 two-photon laser excitations \cite{Santra:2005,Zanon-Willette:2006,Alden:2014}, for ionic systems with a single particle or highly charged systems \cite{Safronova:2014,Yudin:2014}, and molecular clocks \cite{Schiller:2014}. Light-shifts are also important issues in high resolution spectroscopy and tests of theories~\cite{Matveev:2013}.

 In order to reduce shifts and broadening due to inhomogeneous excitation conditions or shifts that are the result of the clock laser excitation itself, sequence of composite excitation pulses based on generalization of the Ramsey scheme \cite{Ramsey:1950} to more complex ones including additional intermediate pulses with suitably selected frequency and phase steps have been introduced.
These new techniques are denoted as Hyper-Ramsey HR schemes~\cite{Yudin:2010} or Generalized Hyper-Ramsey GHR resonances~\cite{Zanon:2015}, and were recently proposed for bosonic three-level systems in Doppler-recoil free configurations using optical stimulated Raman transitions~\cite{Zanon:2014}. They have been demonstrated experimentally for the ultra-narrow electric octupole transition in the $^{171}$Yb$^{+}$ clock \cite{Huntemann:2012}.

As an additional issue for the operation of a very stable optical clock is the stability of the laser probing the clock transition, error signals based on HR schemes have thus been successfully tested to control or cancel probe laser drifts in a single ion clock \cite{Huntemann:2016} and in optical lattice trap with bosons \cite{Hobson:2016}. A precise lock of the probe on the clock transition can be achieved by phase modulation instead of stepwise frequency depending on specific experimental conditions because it is less sensitive to asymmetry in the lineshape. In Ramsey-type schemes \cite{Ramsey:1950}, an error signal correcting the probe drift is obtained by alternately applying phase steps of $\pm\pi/2$ to one of the excitation pulses while keeping the excitation frequency fixed \cite{Ramsey:1951,Morinaga:1989,Letchumanan:2006}.

In this work we propose a very efficient and ultra-stable laser frequency stabilization scheme with composite pulses which completely eliminates the light-shift arising from the laser probe itself while being strongly insensitive to associated large laser pulse area variations. The error signal within the novel scheme is derived from a  control of the probe laser phase: a properly chosen sequence of phase-steps is introduced while applying the GHR clock resonance interrogation in order to realize a robust (and flexible) error signal. We are able to decouple the unperturbed frequency measurement from the laser intensity by producing a laser stabilization of the probe laser monitoring the atomic response to a phase rather than a frequency modulation. We verify the stability of the new locking scheme by simulations including potential variations of the laser power during the clock operation. The main physics behind our error correction scheme, based on ad-hoc designed laser probe pulse sequence and detection, is the symmetry of the atomic response functions with respect to the unshifted resonance frequency. By combining properly chosen symmetrical/antisymmetrical response functions into the error signal it is possible to recover the unperturbed clock frequency.\\

\begin{figure}[t!!]
\centering%
\resizebox{8.0cm}{!}{\includegraphics[angle=-90]{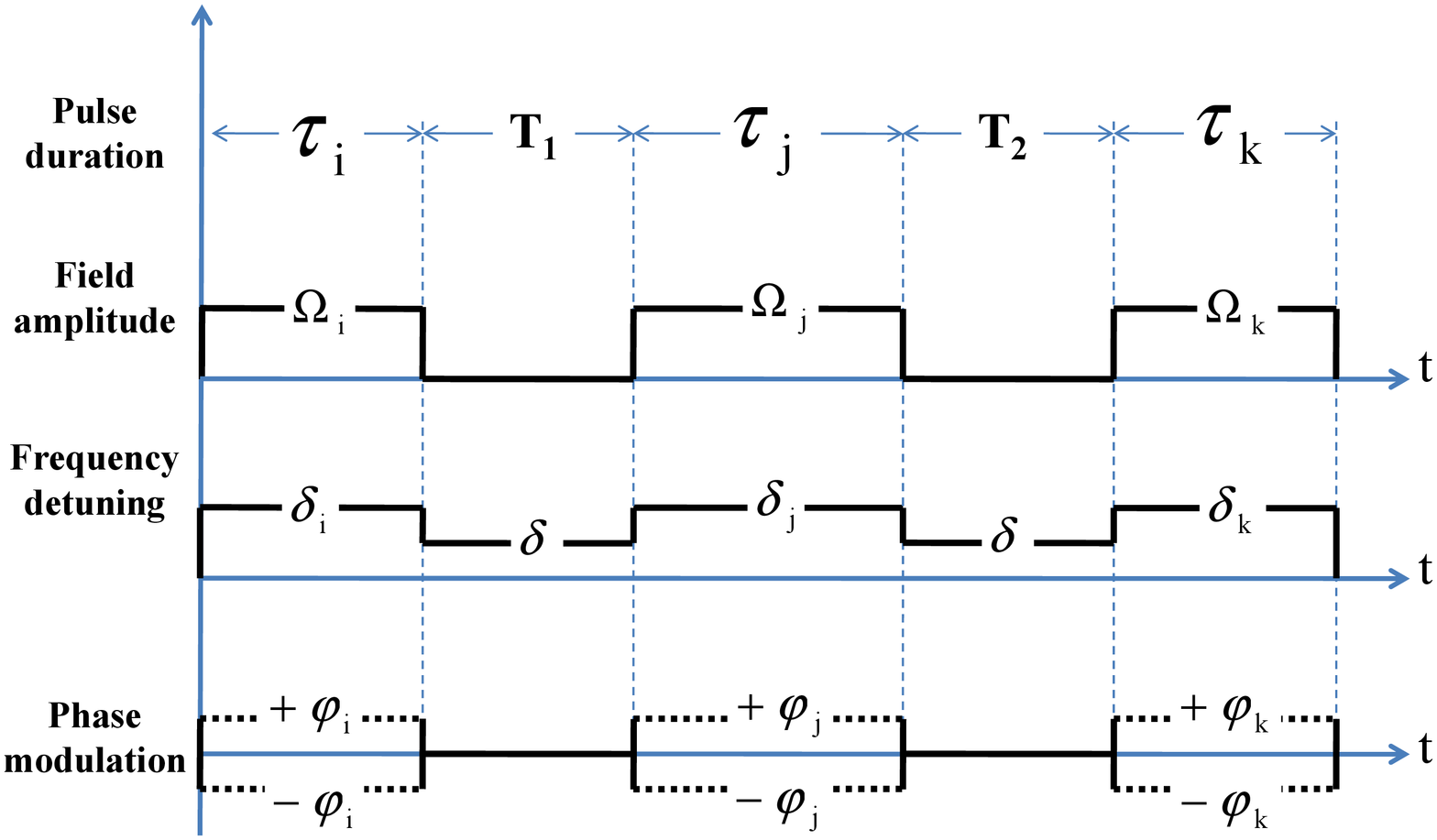}}
\caption{(color online). Time sequence of a composite three-pulse Ramsey interrogation. Parameters of the composite pulse sequence are detuning $\delta_{l}$, complex field amplitude $\Omega_{l}e^{i\varphi_{l}}$, pulse duration $\tau_{l}$ where $l=\textup{i,j,k}$ and the two different free evolution times T$_{1}$ and T$_{2}$ between pulses. Phase step modulations $\varphi_{l}$ of the laser fields are applied to generate a specific and robust error signal.}
\label{figure-1}
\end{figure}

\indent Consider a two-level system of the $|g\rangle,|e\rangle$ clock states which is driven by a sequence of three laser pulses, as schematized in Fig.~\ref{figure-1}.
The basic elements of these sequences with composite pulses are presented in Tab.~\ref{protocol-table-1}. Each sequence is determined by the pulse areas $\theta_\textup{i},\theta_\textup{j},\theta_\textup{k}$ and free evolution times $T_1,T_2$. A laser field phase sequence $\varphi_l$ $(l=\textup{i,\,j,\,k})$ is used by us as a novel control parameter. Various types of GHR sequences of composites pulses with different phase-step modulations are possible. By modifying their characteristics it is possible to produce sequences appropriate for very good control and correction in different experimental configurations. The present work introduces the phase steps into the probing schemes denoted as GHR$_1$ and GHR$_2$ of Table based on the analysis of~\cite{Zanon:2015}. The Table bottom line reports the values used for those phase-steps interrogation protocols. Our best results are obtained when a phase step $\varphi_\textup{j}$ is applied only to the intermediate probe pulse, and we will denote this specific protocol as GHR($\varphi_\textup{j}$). Equivalent results are obtained by introducing the phase step to either GHR$_1$ or GHR$_2$, but all the reported numerical results are based on the first one. For comparison we will examine also a phase-step modulation for the initial HR protocol experimentally used in ion clocks ~\cite{Huntemann:2012,Huntemann:2016}.

\begin{table}[t!!]
\centering%
\caption{Composite pulse sequences. A sequence is made of three laser pulses labeled by their effective areas $\theta_l$, $l=\textup{i,\,j,\,k}$, and one free evolution time $T_1$ or $T_2$. Rabi frequency is $\Omega=\pi/2\tau$ corresponding to initial and final $\pi/2$ pulses. Phase-steps protocols are given for HR and GHR sequences.}
\label{protocol-table-1}
\begin{tabular}{ccccccc}
\hline
\hline
sequences & parameters & $\theta_{\textup{i}}$ & $\delta,T_{1}$  & $\theta_{\textup{j}}$ & $\delta,T_{2}$  & $\theta_{\textup{k}}$  \\
\hline
                 & $\tau_{l}$    & $\tau$      & $T$       & $2\tau$     & $0$  & $\tau$   \\
    \text{GHR$_{1}$}                & $\Omega_{l}e^{\imath\varphi_{l}}$  & $\Omega e^{\imath\varphi_{\textup{i}}}$ & $0$       & $\Omega e^{\imath\varphi_{\textup{j}}}$ &  $0$ & $\Omega              e^{\imath\varphi_{\textup{k}}}$   \\
                 & $\delta_{l}$  & $\delta+\Delta_{\textup{i}}$    & $\delta$  & $\delta+\Delta_{\textup{j}}$    &  $0$ & $\delta+\Delta_{\textup{k}}$   \\ \\

                 & $\tau_{l}$    & $\tau$      & $0$     & $2\tau$  & $T$  & $\tau$   \\
    \text{GHR$_{2}$}                 & $\Omega_{l}e^{\imath\varphi_{l}}$  & $\Omega e^{\imath\varphi_{\textup{i}}}$ & $0$  & $\Omega e^{\imath\varphi_{\textup{j}}}$ &  $0$ & $\Omega e^{\imath\varphi_{\textup{k}}}$   \\
                 & $\delta_{l}$  & $\delta+\Delta_{\textup{i}}$    & $0$  & $\delta+\Delta_{\textup{j}}$    &  $\delta$ & $\delta+\Delta_{\textup{k}}$   \\ \\ \\
\hline
\text{phase-steps}   &              & $\varphi_{\textup{i}}$ & $$  & $\varphi_{\textup{j}}$ & $$  & $\varphi_{\textup{k}}$  \\
\hline \\
             HR     & & $\pm\frac{\pi}{2}$ &  &  $\pi$               &  & $0$  \\\\

             GHR($\varphi_{l}$)   & &  $0$              &  & $\pm\varphi_{l}$    &  & $0$  \\\\

\hline
\hline
\end{tabular}
\end{table}

\indent Recalling the GHR treatment~\cite{Zanon:2015}, the atomic superposition of the clock following a pulse excitation is
\begin{equation}
|\Psi(\theta_{l})\rangle=c_g(\theta_{l})|g\rangle+c_e(\theta_{l})|e\rangle.
\end{equation}
characterized by the pulse area $\theta_{l}=\omega_{l}\tau_{l}/2$ with generalized Rabi frequency $\omega_{l}=\sqrt{\delta_{l}^{2}+\Omega_{l}^{2}}$.
If during the pulses, light-shifts from off-resonant states are present, a laser frequency step $\Delta_{l}$ is applied to the unperturbed $\delta$ probe detuning in order to rectify the anticipated shift, thus requiring a redefinition of frequency detunings as $\delta_{l}=\delta+\Delta_{l}$.
Using the solution of the Schr\"odinger's equation, we write for the $c_{g,e}(\theta_{l})$ ($l=\textup{i,j,k}$) transition amplitudes
\begin{equation}
\begin{split}
\left(%
\begin{array}{c}
c_{g}(\theta_{l})
\\
c_{e}(\theta_{l})
\\
\end{array}%
\right)
=\chi(\theta_{l})\cdot \textup{M}(\theta_{l})\cdot
\left(%
\begin{array}{c}
c_{g}(0) \\
c_{e}(0) \\
\end{array}%
\right)
\end{split}
\end{equation}
including a phase factor of the form $\chi(\theta_{l})=\exp\left[-i\delta_{l}\frac{\tau_{l}}{2}\right]$.
The wave-function evolution driven by the pulse area $\theta_{l}$ is given by a complex 2x2 spinor interaction matrix~\cite{Zanon:2015}:
\begin{equation}
\begin{split}
\textup{M}(\theta_{l})&=\left(
\begin{array}{cc}
\textup{M}_{+}(\theta_{l}) &e^{\imath\varphi_{l}}\textup{M}_{\dagger}(\theta_{l}) \\
e^{-\imath\varphi_{l}}\textup{M}_{\dagger}(\theta_{l}) & \textup{M}_{-}(\theta_{l}) \\
\end{array}%
\right)\\
&=\left(%
\begin{array}{cc}
\cos(\theta_{l})+i\frac{\delta_{l}}{\omega_{l}}\sin(\theta_{l})&-ie^{i\varphi_{l}}\frac{\Omega_{l}}{\omega_{l}}\sin(\theta_{l}) \\
-ie^{-i\varphi_{l}}\frac{\Omega_{l}}{\omega_{l}}\sin(\theta_{l})&\cos(\theta_{l})-i\frac{\delta_{l}}{\omega_{l}}\sin(\theta_{l}) \\
\end{array}
\right).
\end{split}
\label{Matrix}
\end{equation}
The overall evolution by the pulsed excitations is given by a product of different matrices M$(\theta_{l})$ for the laser pulses and M$(\delta T_1)$ (M$(\delta T_2)$) for the free evolutions without laser light during time $T_{1}$ ( $T_{2}$), respectively.
Applying the above matrix to the $c_{g}(0)=1,c_{e}(0)=0$ initial conditions, the final expression of the complex amplitude is computed, leading to the transition probability $P_{|g\rangle\mapsto|e\rangle}$ expressed as
\begin{equation}
\begin{split}
P_{|g\rangle\mapsto|e\rangle}=|\langle e|\textup{M}(\theta_{\textup{k}})\textup{M}(\delta T_{2})\textup{M}(\theta_{\textup{j}})\textup{M}(\delta T_{1})\textup{M}(\theta_{\textup{i}})|g\rangle|^{2}.
\label{Generalized-Hyper-Ramsey-transition}
\end{split}
\end{equation}
The measured spectroscopic signal is the fraction of the atomic population $P_{|g\rangle\mapsto|e\rangle}$ following the application of different phase-step modulations of the laser pulses within Eq.~\eqref{Generalized-Hyper-Ramsey-transition}. The dispersive-shaped error signal $\Delta E$ is obtained by taking the difference between two spectroscopic signals with phase-steps having opposite sign as:
\begin{subequations}
\begin{align}
\Delta E_{\textup{GHR}(\varphi_{\textup{j}})}=P(0,+\varphi_{\textup{j}},0)_{|g\rangle\mapsto|e\rangle}-P(0,-\varphi_{\textup{j}},0)_{|g\rangle\mapsto|e\rangle},\label{Error-signal-GHR}\\
\Delta E_{\textup{HR}}=P(\pi/2,\pi,0)_{|g\rangle\mapsto|e\rangle}-P(-\pi/2,\pi,0)_{|g\rangle\mapsto|e\rangle},\label{Error-signal-HR}
\end{align}
\label{Error-signal}
\end{subequations}
\noindent where we reference to the two phase-steps protocols introduced in Tab.~\ref{protocol-table-1}.
An imperfect $\delta \nu$ light-shift compensation leads to the $\Delta E=0$ clock lock condition for $\delta=\delta\nu\ne0$. Instead the correct target is to satisfy the lock condition with $\delta=0$.\\
\begin{figure}[t!!]
\resizebox{9.0cm}{!}{\includegraphics[angle=0]{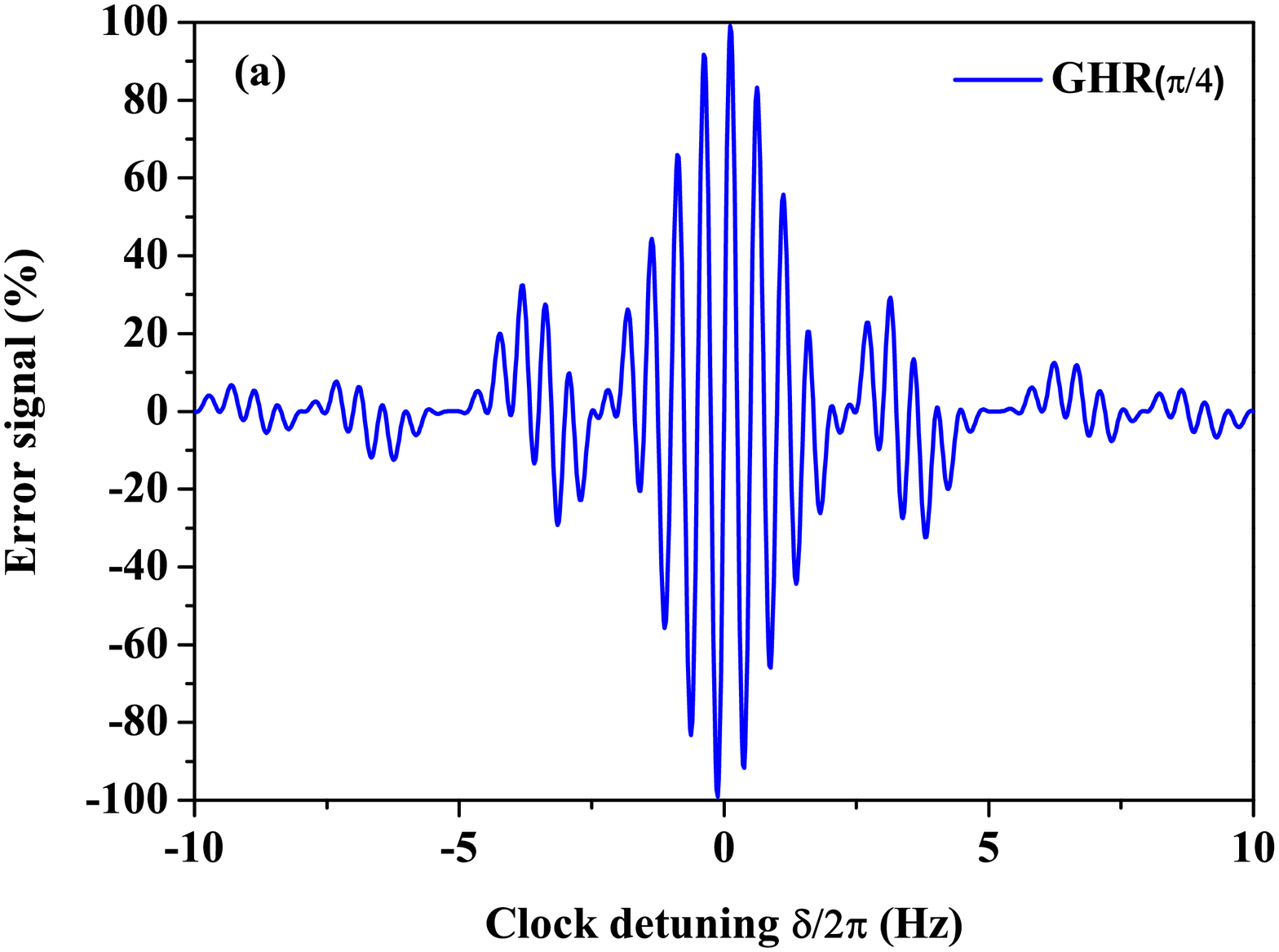}}
\resizebox{9.0cm}{!}{\includegraphics[angle=0]{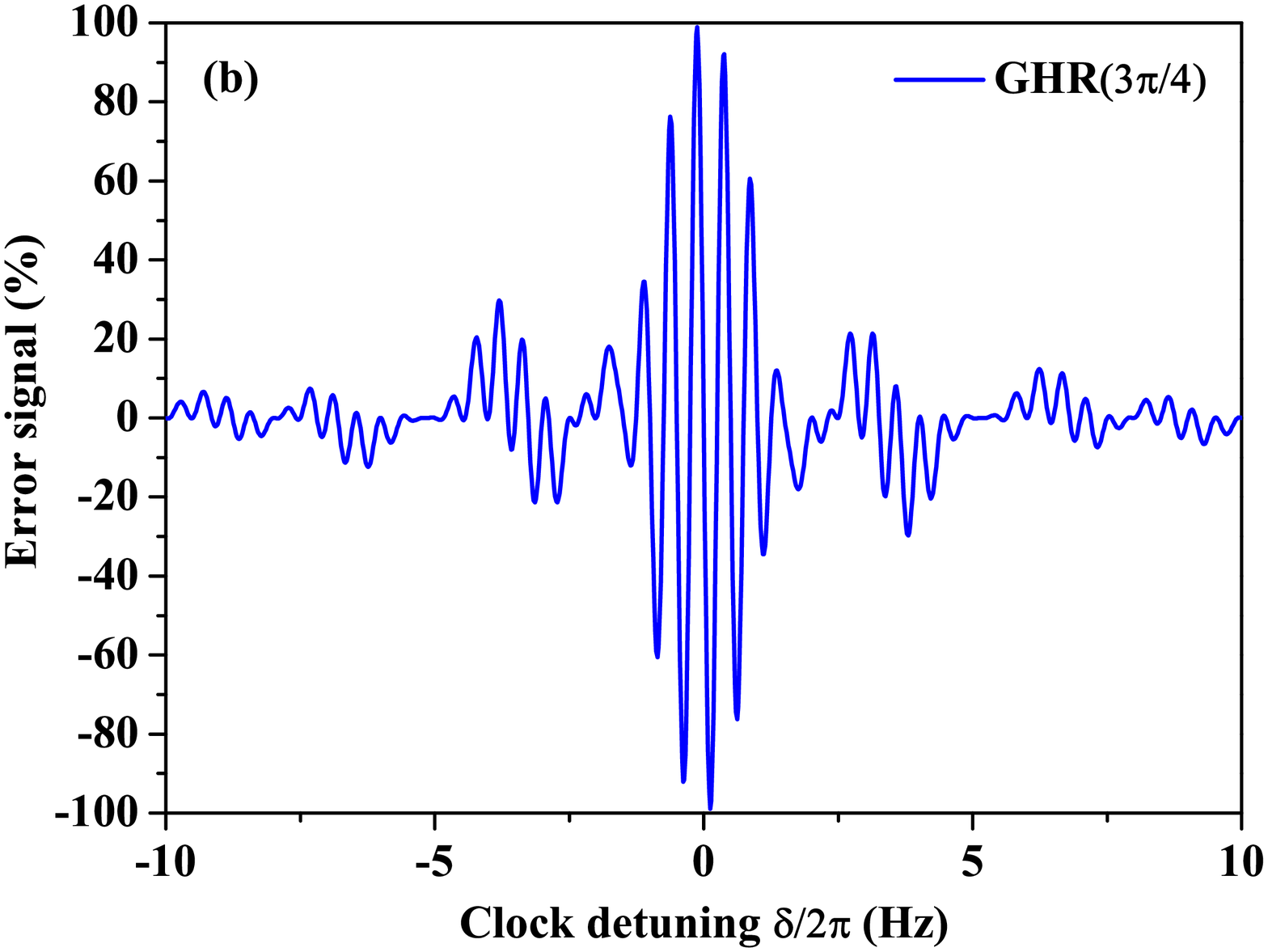}}
\caption{(color online). Generalized Hyper-Ramsey dispersive error signals versus the frequency clock detuning $\delta$ with $\Delta=0$. (a) $\Delta E_{\textup{GHR}(\pi/4)}$ and (b) $\Delta E_{\textup{GHR}(3\pi/4)}$ are calculated with Eq.~\eqref{Generalized-Hyper-Ramsey-transition} and Eq.~\eqref{Error-signal-GHR}. The pulse duration is $\tau=0.1875~$s, the Rabi frequency is fixed to $\Omega=\pi/2\tau$ and the free evolution time is $T=2~$s.}
\label{figure-2}
\end{figure}
To drastically suppress the probe induced shifts over very large uncompensated offsets, we use the GHR phase-step modulation from Tab.~\ref{protocol-table-1} where only the phase of the intermediate pulse experiences a step by $\pm\pi/4$ or $\pm3\pi/4$.
Making use of Eq.~\eqref{Error-signal-GHR}, we report the synthesized error signals with the GHR($\varphi_{\textup{j}}$) protocols for $\varphi_{\textup{j}}=\pm\pi/4$ in Fig.~\ref{figure-2}(a) and for $\varphi_{\textup{j}}=\pm3\pi/4$ in Fig.~\ref{figure-2}(b).\\
\indent We derive a simple analytical formula for the laser error signal, that provides a direct physical insight and a straightforward determination of the best parameters for the shift compensation. The full expression of the GHR error signal is found to be:
\begin{equation}
\begin{split}
\Delta E_{\textup{GHR}(\varphi_{\textup{j}})}=&\frac{16\Omega^{6}}{\omega^{6}}H_{\delta,\Delta}\sqrt{1+\tan^{2}\Theta}\times\cos(\delta T/2-\Theta)\nonumber\\
&\times\sin\varphi_{j}\sin^{2}\theta\sin(\delta T/2).
\label{Full-GHR-error-signal}
\end{split}
\end{equation}
where $\tan\Theta=G_{\delta,\Delta}/H_{\delta,\Delta}$. Functions $G_{\delta,\Delta}$ and $H_{\delta,\Delta}$ are:
\begin{equation}
\small{
\begin{split}
G_{\delta,\Delta}=&\frac{(\delta+\Delta)\omega}{\Omega^{2}}\left[\frac{(\delta+\Delta)^{2}}{\Omega^{2}}+\cos2\theta\right]\left(2\cos\theta+\cos3\theta\right)\sin\theta,\\
H_{\delta,\Delta}=&\cos^{2}\theta\left[4\cos\varphi_{j}\sin^{2}\theta\left(\frac{(\delta+\Delta)^{2}}{\Omega^{2}}+\cos^{2}\theta\right)\right.\\
&\left.-\frac{\omega^{2}}{\Omega^{2}}\left(\frac{(\delta+\Delta)^{2}}{\Omega^{2}}+\cos2\theta\right)\left(\cos2\theta-\frac{2(\delta+\Delta)^{2}}{\omega^{2}}\sin^{2}\theta\right)\right].
\end{split}}
\end{equation}
A good approximation can be made around the central dispersive feature when $\delta,\Delta\ll\Omega$, with $\Omega\tau\sim\pi/2$, leading to an analytical expression of $\Delta E_{\textup{GHR}(\varphi_{\textup{j}})}$ for the GHR protocol as:
\begin{equation}
\begin{split}
\Delta E_{\textup{GHR}(\varphi_{\textup{j}})}\approx&~4\left( \frac{\Omega}{\omega}\right)^{6}\left[\left( \frac{\delta+\Delta}{\Omega}\right) ^{2}+\cos^{2}\theta\right]\sin^{2}2\theta\sin^{2}\theta\\
&\times\sin(2\varphi_{\textup{j}})\sin(\delta T).
\label{GHR-error-signal-approximation}
\end{split}
\end{equation}
For instance, with parameter values of Fig.~\ref{figure-2}, the fractional difference between results of Eq.~\eqref{GHR-error-signal-approximation} and Eq.~\eqref{Full-GHR-error-signal} is less than $10^{-3}$ in the range $-0.1$~Hz $\leqslant\delta/2\pi\leqslant0.1$~Hz for an uncompensated residual shift $\Delta/2\pi=0.1$~Hz.

Eq.~\eqref{GHR-error-signal-approximation} emphasizes the robustness of our correction scheme. For the optimum case $\varphi_{\textup{j}}=\pm\pi/4$ or $\varphi_{\textup{j}}=\pm3\pi/4$, a rapid analysis of the unperturbed $\sin(\delta T)$ dependence appearing in Eq.~\eqref{GHR-error-signal-approximation} tells us that the frequency lock point does not suffer at all from residual offset errors. The error signal contrast is also nearly immune to strong $\pm 10\%$ variations of the laser field amplitude because of the $(\Omega/\omega)^{6}\sin^{4}2\theta$ dependence and the required $\theta\sim\pi/4$ pulse area. By operating also near the minima/maxima of the $\sin2\varphi_{\textup{j}}$ term, a large insensitivity to the exact phase values is obtained.
\begin{figure}[t!!]
\resizebox{8.5cm}{!}{\includegraphics[angle=0]{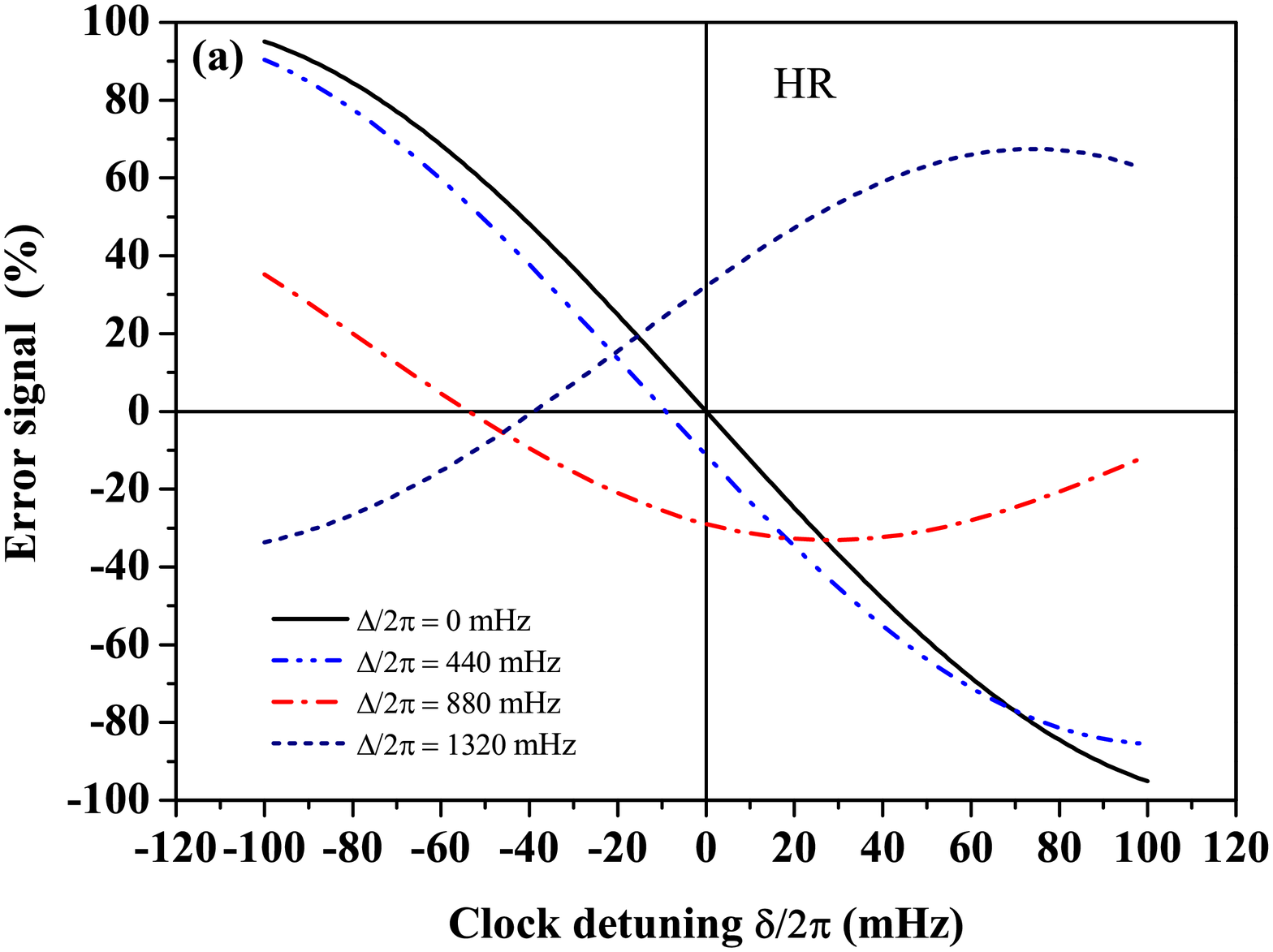}}
\resizebox{8.5cm}{!}{\includegraphics[angle=0]{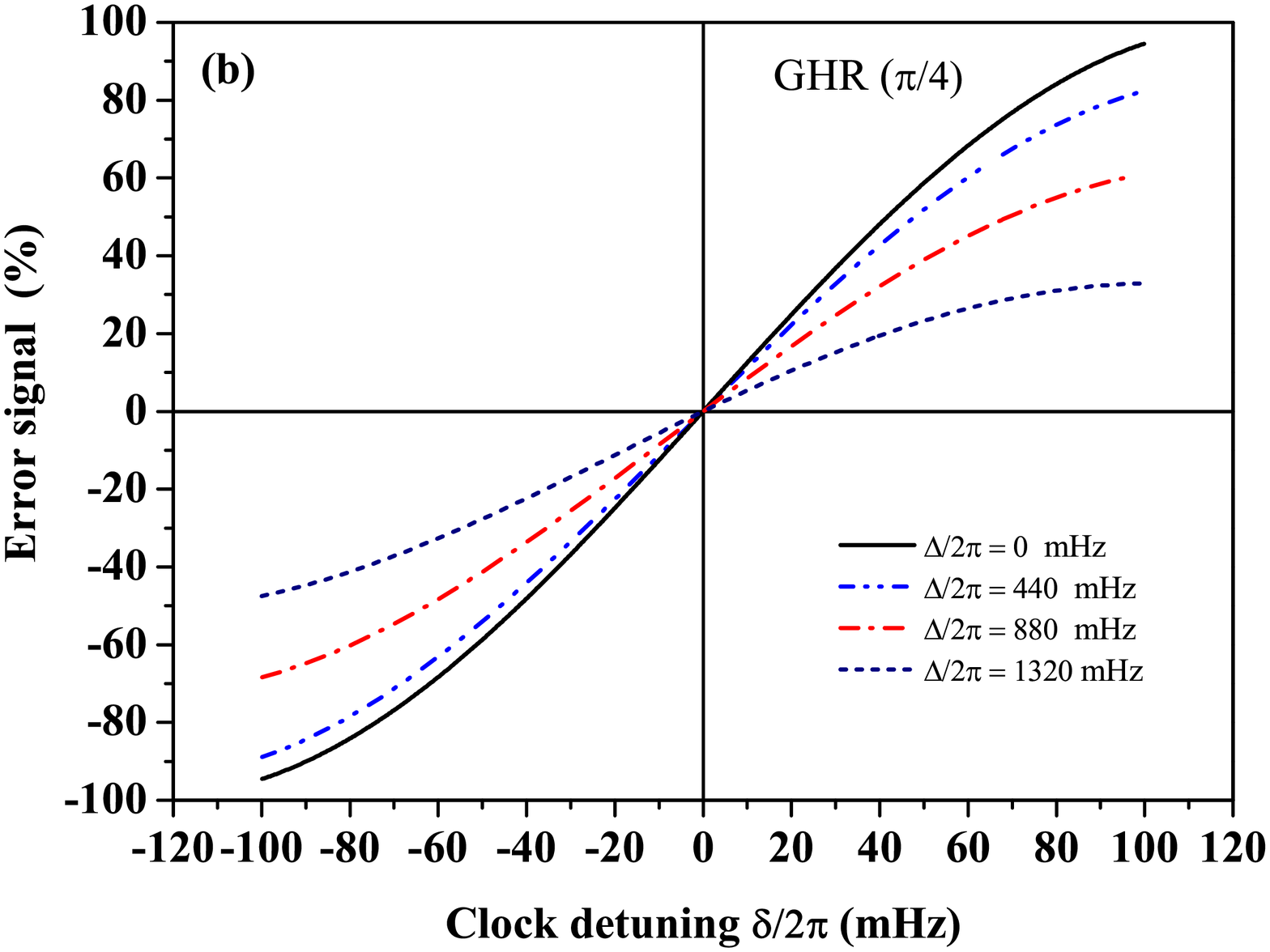}}
\resizebox{8.5cm}{!}{\includegraphics[angle=0]{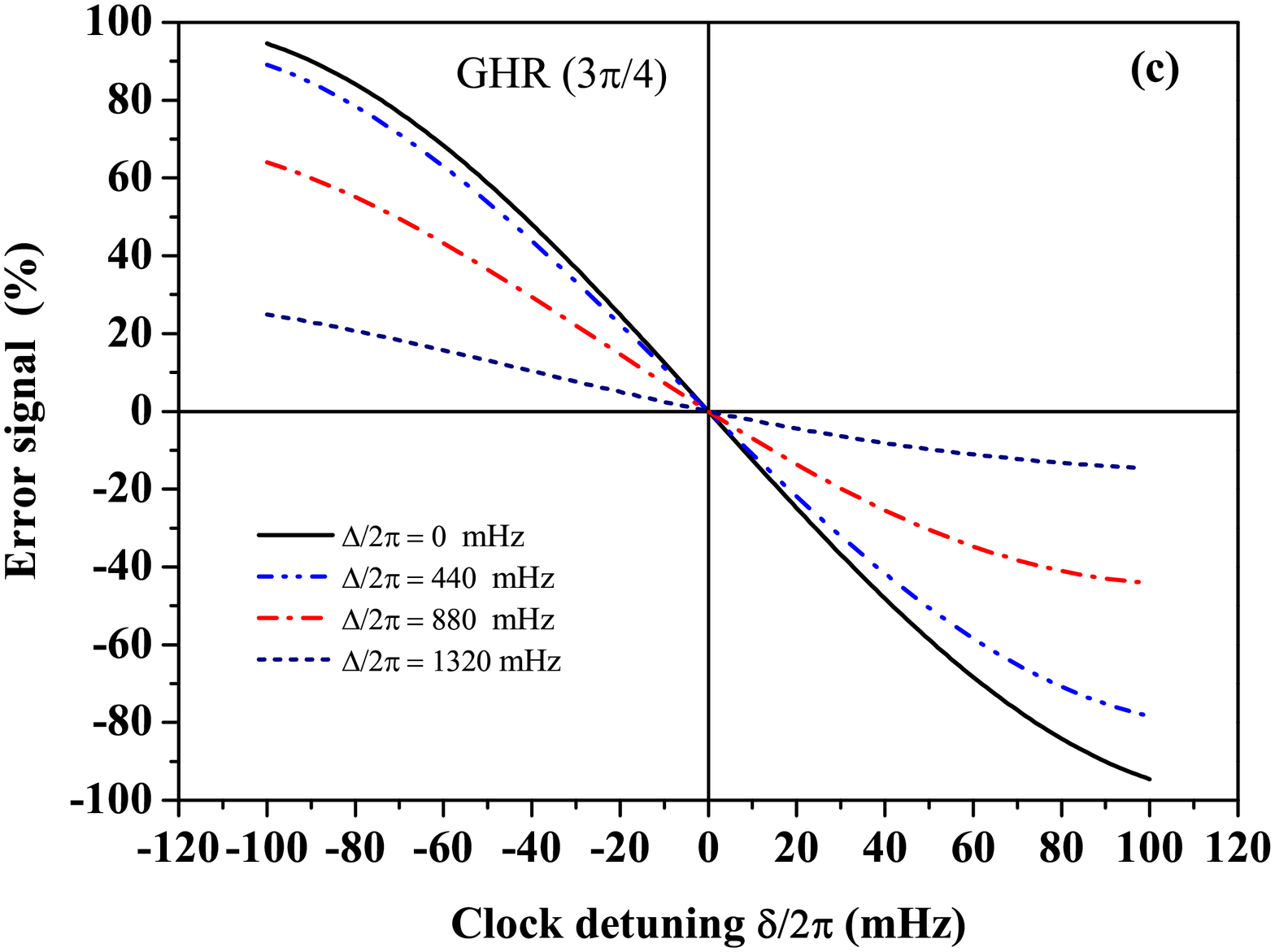}}
\caption{(color online). Error signal $\Delta E$ vs $\delta$ in presence of residual light-shift perturbations $\Delta\neq0$ (we use $\Delta_{\textup{i,j,k}}=\Delta$). In (a), (b) and (c) three phase-steps protocols, $\Delta E_{\textup{HR}}$, $\Delta E_{\textup{GHR}(\pi/4)}$ and $\Delta E_{\textup{GHR}(3\pi/4)}$ respectively. Same conditions as in Fig.~\ref{figure-2}.}
\label{figure-3}
\end{figure}
The $\theta$ pulse area dependencies in Eqs. \eqref{Full-GHR-error-signal} and \eqref{GHR-error-signal-approximation}, as for the standard Ramsey pulses, allow a compensation for variations in the laser pulse parameters. In practice a not-constant Rabi frequency during rise and fall times may limit the validity of such results, and only numerical results for the specific experimental parameters allow to test the error signal stability.
\begin{figure}[t!!]
\resizebox{9.0cm}{!}{\includegraphics[angle=0]{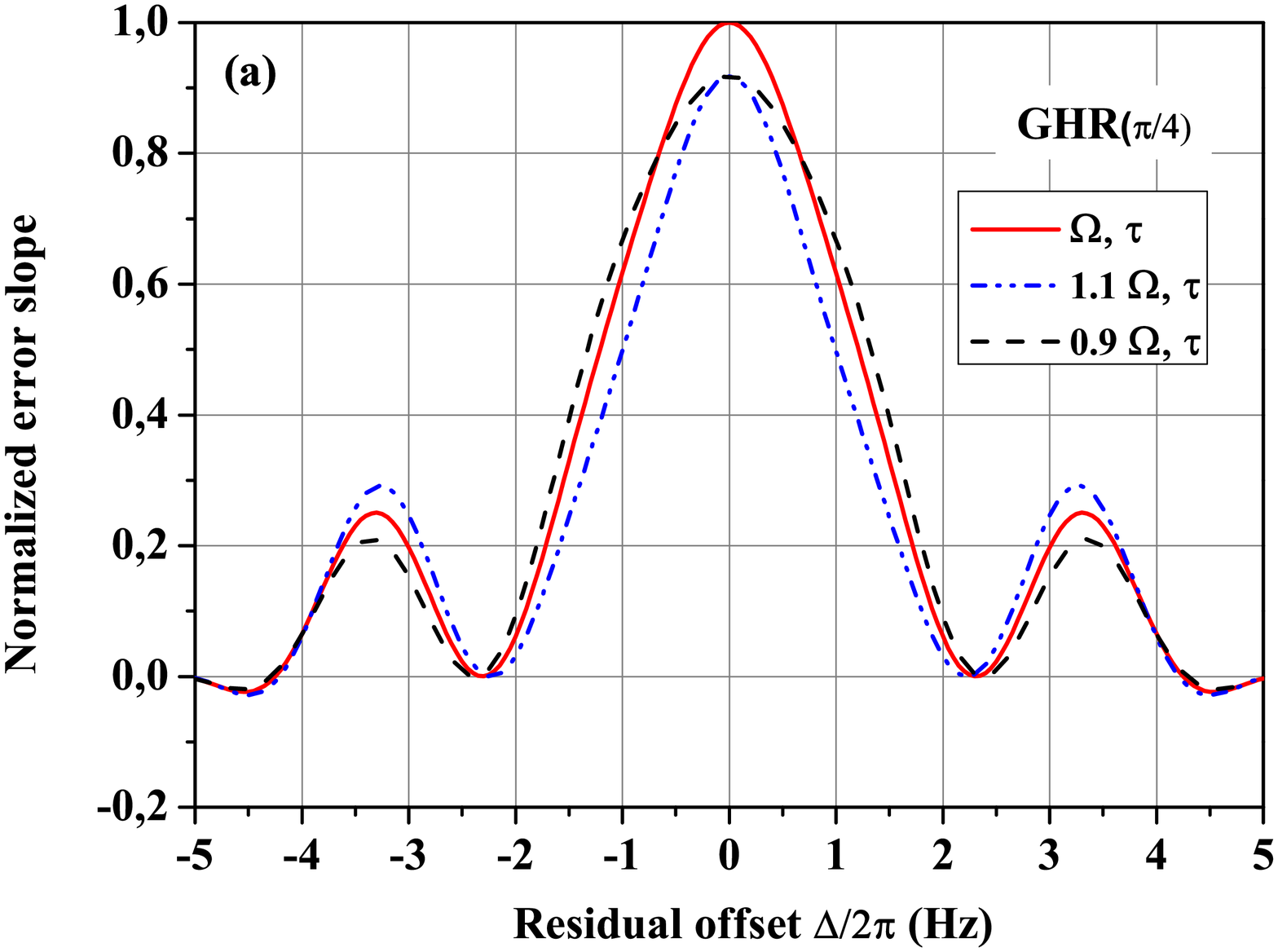}}
\resizebox{9.0cm}{!}{\includegraphics[angle=0]{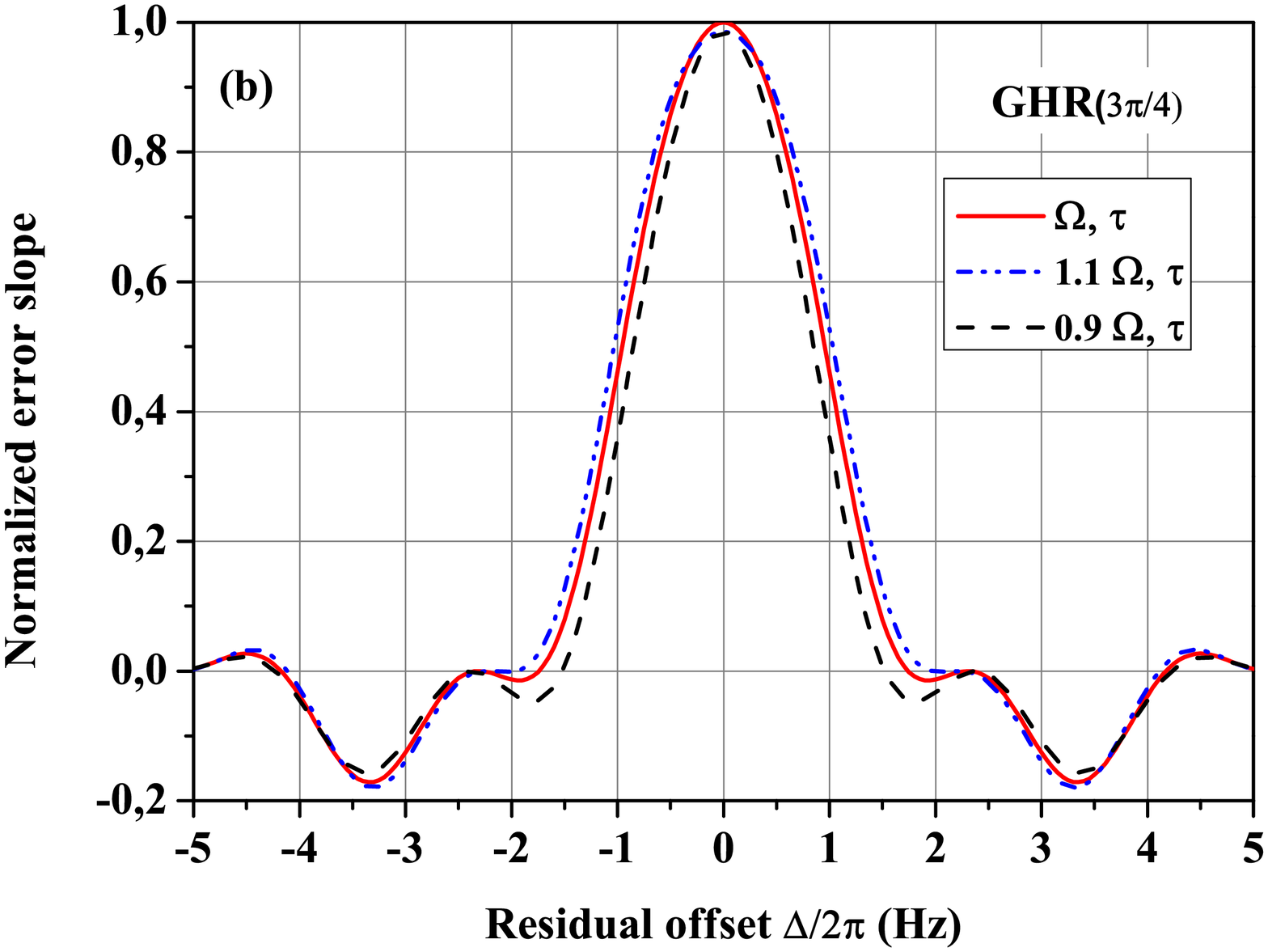}}
\caption{(color online). Normalized slopes of error signals (a) GHR($\pi/4$) and (b) GHR($3\pi/4$) versus the uncompensated residual offset. The slope predicted by Eq.~\eqref{Full-GHR-error-signal} is plotted for Rabi frequency increased or decreased by $10\%$ with a fixed pulse duration $\tau$. All others conditions as in Fig.~\ref{figure-2}.}
\label{figure-4}
\end{figure}
We compare now the GHR phase-step protocol to the HR protocol of Tab.~\ref{protocol-table-1} based on the error signal generated by Eq.~\eqref{Error-signal-HR}.
By applying this last servo-locking scheme, the linear dependence of the $\Delta E_{\textup{HR}}$ error signal on small variations of the light-shifts is nearly totally removed except for a residual cubic correction \cite{Yudin:2010,Zanon:2015}. Fig.~\ref{figure-3}(a) reports the response of the HR frequency lock point to some residual offset ac Stark-shifts $\Delta$ and Fig.~\ref{figure-3}(b) and (c) compare it, under the same conditions, to the GHR scheme result for the two possible phase-step modulations. Contrary to the HR protocol, in presence of large residual offsets due to imperfect compensation of light-shift, the $\Delta E_{\textup{GHR}(\pi/4)}$ and $\Delta E_{\textup{GHR}(3\pi/4)}$ error signals have a perfectly protected lock point which experiences a rotation around the unperturbed frequency, i.e., $\delta=0$, with a slightly reduced slope and a small distortion.
Note that a reduced slope results in a clock stability reduction as the frequency stability scales as the inverse of the slope. The error signal behavior versus the residual shift for both GHR protocols are shown in Fig.~\ref{figure-4}(a) and(b) for ten percent variations of $\Omega$. An $10\%$ error on the laser field amplitude modifies the slope value but does not degrade the locking range, the range of $\Delta$ where the slope is not null. This frequency-locking bandwidth is governed by $\tau$ and it increases for smaller $\tau$ values.
\begin{figure}[t!!]
\resizebox{9.0cm}{!}{\includegraphics[angle=0]{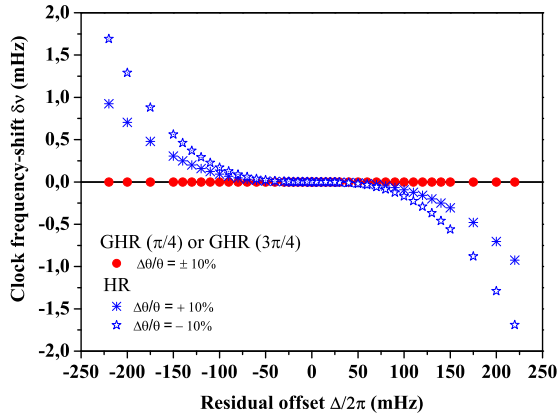}}
\caption{(color online). Sensitivity of the clock frequency-shift $\delta\nu$ for HR and GHR protocols to some pulse areas variation of $\Delta\theta/\theta=\pm10\%$ during laser pulses versus uncompensated residual offsets. All others conditions as in Fig.~\ref{figure-2}.}
\label{figure-5}
\end{figure}
Fig.~\ref{figure-5} compares both schemes, HR and GHR, in terms of the associated clock frequency-shift $\delta\nu$ versus residual uncompensated offset corrections under a relative modification of $\pm10\%$ of all pulse areas. For high residual offsets, the sensitivity of the error signal $\Delta E_{\textup{HR}}$ to laser intensity variation becomes more pronounced as opposed to the $\Delta E_{\textup{GHR}(\pi/4)}$ (or the $\Delta E_{\textup{GHR}(3\pi/4)}$) signal error which is immune as shown by the unperturbed red dots. It is worth to note that the clock frequency shift remains eliminated even in presence of those pulse area variations.
From the previous analysis, both residual uncompensated offsets and pulse area variations are thus directly transferred to a slope reduction of error signals with no change in their frequency lock point. While the HR protocol has a residual uncompensated frequency-shift around 1-2~mHz over a 220~mHz residual offset, the GHR($\pi/4$) or the GHR($3\pi/4$) scheme lead to a total suppression of frequency-shifts induced by laser probing fields opening the access to a fractional level well below $10^{-18}$ for any optical clock.

We note recently that a different method based on modified Hyper-Ramsey spectroscopy with another phase-step modulation scheme has been proposed and successfully demonstrated on bosonic $^{88}$Sr neutral atoms \cite{Hobson:2016}, demonstrating automatic suppression of a sizable $2\times10^{-13}$ probe Stark shift to below $10^{-16}$ even with very large errors in shift compensation.\\

\indent The schemes presented here are well protected against the routinely considered sources of clock errors, as laser intensity variations and residual offsets induced by laser probe fields within an ideal trap environment. However the incredible accuracy level reached by the present (and future) optical clocks may require to examine some additional external error sources, as the atomic population relaxation and the laser linewidth (see very recent work by~\cite{Yudin:2016}). The robustness of all schemes against these additional error sources should be tested, and the present Schr\"odinger equation approach cannot be directly applied. The schemes may thus become unstable and the robustness of the frequency lock point might be lost.
For example, the density-matrix analysis of \cite{Yudin:2016} has demonstrated that, under the condition $\Omega\neq\pi/2\tau$ and in presence of laser induced decoherence, in contrary to the modified Hyper-Ramsey scheme \cite{Hobson:2016}, the GHR($\pi/4$) or GHR($3\pi/4$) phase-step modulation schemes do not exhibit additional parasitic shifts when the residual offsets are perfectly well compensated.\\
\indent In conclusion, we propose a new ultra-stable laser frequency stabilization scheme based on Generalized Hyper-Ramsey spectral resonances with different phase-step modulations that strongly reduce the probe induced frequency shift and produce a strong immunity against laser intensity variations to a high degree of robustness. Compared to the modified Hyper-Ramsey method of \cite{Hobson:2016}, the synthesized error signal is derived from spectra acquired with different phase-steps in the intermediate $\pi$-pulse allowing a complete suppress of light shifts related to the excitation pulses. The application of the present frequency locking technique will really boost metrological performances of all types of quantum clocks. In particular our laser frequency locking technique may have a direct impact on the next generation of quantum clocks with bosonic species for probing completely dipole forbidden atomic transitions suffering from important ac Stark-shifts \cite{Taichenachev:2006,Ovsiannikov:2007,Zanon-Willette:2006,Zanon:2014}.

\indent T. Zanon-Willette deeply acknowledges V.I. Yudin, A.V. Taichenachev for stimulating and challenging discussions, C. Janssen, B. Darquié, J. Ye, M. Glass-Maujean for suggestions and a careful reading of the manuscript.


\begin{thebibliography}{1}
\bibitem{Ludlow:2015} A.D. Ludlow, M.M. Boyd, J. Ye, E. Peik and P.O. Schmidt, Optical atomic clocks, Rev. Mod. Phys. \textbf{87}, 637 (2015).
\bibitem{Derevianko:2011} A. Derevianko, H. Katori, Colloquium: Physics of optical lattice clocks, Rev. Mod. Phys. \textbf{83}, 331 (2011).
\bibitem{Akatsuka:2010} T. Akatsuka, M. Takamoto, and H. Katori, Three-dimensional optical lattice clock with bosonic $^{88}$Sr atoms, Phys. Rev. A \textbf{81}, 023402 (2010).
\bibitem{Ludlow:2011} A.D. Ludlow, N. D. Lemke, J.A. Sherman, C.W. Oates, G. Quéméner, J. von Stecher, and A.M. Rey, Cold-collision-shift cancellation and inelastic scattering in a Yb optical lattice clock, Phys. Rev. Lett. \textbf{84}, 052724 (2011).
\bibitem{Nicholson:2015} T.L. Nicholson, S.L. Campbell, R.B. Hutson, G.E. Marti, B.J. Bloom, R.L. McNally, W. Zhang, M.D. Barrett, M.S. Safronova, G.F. Strouse, W.L. Tew and J. Ye, Systematic evaluation of an atomic clock at 2$\times$10$^{-18}$ total uncertainty, Nature Comm. \textbf{6}, 7896 (2015).
\bibitem{Taichenachev:2006} A.V. Taichenachev, V.I. Yudin, C.W. Oates, C.W. Hoyt, Z.W. Barber and L. Hollberg, Magnetic field-induced spectroscopy of forbidden optical transitions with application to lattice-based optical atomic clocks, Phys. Rev. Lett. \textbf{96}, 083001 (2006).
\bibitem{Barber:2006} Z. Barber, C. Hoyt, C. Oates, L. Hollberg, A. Taichenachev and V. Yudin, Direct excitation of the forbidden clock transition in neutral $^{174}$Yb atoms confined to an optical lattice, Phys. Rev. Lett. \textbf{96}, 083002 (2006).
\bibitem{Baillard:2007} X. Baillard, M. Fouch\'{e}, R. Le Targat, P.G. Westergaard, A. Lecallier, Y. Le Coq, G.D. Rovera, S. Bize, and P. Lemonde, Accuracy evaluation of an optical lattice clock with bosonic atoms, Opt. Lett. \textbf{32}, 1812 (2007).
\bibitem{Kulosa:2015} A.P. Kulosa, D. Fim, K.H. Zipfel, S. Rühmann, S. Sauer, N. Jha, K. Gibble, W. Ertmer, E.M. Rasel, M.S. Safronova, U.I. Safronova, S.G. Porsev, Towards a Mg Lattice Clock: Observation of the $^{1}S_{0}$-$^{3}P_{0}$ Transition and Determination of the Magic Wavelength, Phys. Rev. Lett. \textbf{115}, 240801 (2015).
\bibitem{Ovsiannikov:2007} V.D. Ovsiannikov, V.G. Pal'chikov, A.V. Taichenachev, V.I. Yudin, H. Katori, and M. Takamoto, Magic-wave-induced $^1$S$_0$-$^3$P$_0$ transition in even isotopes of alkaline-earth-metal-like atoms, Phys. Rev. A,  \textbf{75}, 020501(R), (2007).
\bibitem{Santra:2005} R. Santra, E. Arimondo, T. Ido, C.H. Greene, and J. Ye, High-Accuracy optical clock via three-level coherence in neutral bosonic $^{88}$Sr, Phys. Rev. Lett. \textbf{94}, 173002 (2005).
\bibitem{Zanon-Willette:2006} T. Zanon-Willette, A.D. Ludlow, S. Blatt, M.M. Boyd, E. Arimondo, J. Ye, Cancellation of Stark shifts in optical lattice clocks by use of pulsed Raman and electromagnetically induced transparency techniques, Phys. Rev. Lett. \textbf{97}, 233001 (2006).
\bibitem{Alden:2014} E.A. Alden, K.R. Moore, and A.E. Leanhardt, Two-photon $E1-M1$ optical clock, Phys. Rev. A \textbf{90}, 012523 (2014).
\bibitem{Safronova:2014} M.S. Safronova, V.A. Dzuba, V.V. Flambaum, U.I. Safronova, S.G. Porsev and M.G. Kozlov, Highly charged ions for atomic clocks, quantum information, and search for $\alpha$ variation, Phys. Rev. Lett. \textbf{113}, 030801 (2014).
\bibitem{Yudin:2014} V.I. Yudin, A.V. Taichenachev and A. Derevianko, Magnetic dipole transitions in highly charged ions as a basis of ultraprecise optical clocks, Phys. Rev. Lett. \textbf{113}, 233003 (2014).
\bibitem{Schiller:2014} S. Schiller, D. Bakalov and V.I. Korobov, Simplest molecules as candidates for precise optical clocks, Phys. Rev. Lett. \textbf{113}, 023004 (2014).
\bibitem{Matveev:2013} A. Matveev, C.G. Parthey, K. Predehl, J. Alnis, A. Beyer, R. Holzwarth, T. Udem, T. Wilken, N. Kolachevsky, M. Abgrall, D. Rovera, C. Salomon, P. Laurent, G. Grosche, O. Terra, T. Legero, H. Schnatz, S. Weyers, B. Altschul, and T.W. H\"ansch, Precision measurement of the hydrogen 1S-2S frequency via a 920-km fiber link, Phys. Rev. Lett. \textbf{110}, 230801 (2013).
\bibitem{Ramsey:1950} N.F. Ramsey, A molecular beam resonance method with separated oscillating fields, Phys. Rev. \textbf{78}, 695 (1950).
\bibitem{Yudin:2010} V.I. Yudin, A.V. Taichenachev, C.W. Oates, Z.W. Barber, N.D. Lemke, A.D. Ludlow, U. Sterr, Ch. Lisdat and F. Riehle, Hyper-Ramsey spectroscopy of optical clock transitions, Phys. Rev. A \textbf{82}, 011804(R) (2010).
\bibitem{Zanon:2015} T. Zanon-Willette, V.I. Yudin, A.V. Taichenachev, Generalized hyper-Ramsey resonance with separated oscillating fields, Phys. Rev. A \textbf{92}, 023416 (2015).
\bibitem{Zanon:2014} T. Zanon-Willette, S. Almonacil, E. de Clercq, A.D. Ludlow and E. Arimondo, Quantum engineering of atomic phase shifts in optical clocks, Phys. Rev. A \textbf{90}, 053427 (2014).
\bibitem{Huntemann:2012} N. Huntemann, B. Lipphardt, M. Okhapkin, Chr. Tamm, E. Peik, A.V. Taichenachev and V.I. Yudin, Generalized Ramsey excitation scheme with suppressed light shift, Phys. Rev. Lett. \textbf{109}, 213002 (2012).
\bibitem{Huntemann:2016} N. Huntemann, C. Sanner, B. Lipphardt, Chr. Tamm and E. Peik, Single-Ion Atomic Clock with $3\times10^{-18}$ Systematic Uncertainty, Phys. Rev. Lett. \textbf{116}, 063001 (2016).
\bibitem{Hobson:2016} R. Hobson, W. Bowden, S.A. King, P.E.G. Baird, I.R. Hill, P. Gill, Modified hyper-Ramsey methods for the elimination of probe shifts in optical clocks, Phys. Rev. A \textbf{93}, 010501(R) (2016).
\bibitem{Ramsey:1951} N.F. Ramsey and H.B. Silsbee, Phase shifts in the molecular beam method of separated oscillating fields, Phys. Rev. \textbf{84}, 506 (1951).
\bibitem{Morinaga:1989} A. Morinaga, F. Riehle, J. Ishikawa, and J. Helmcke, A Ca optical frequency standard:
frequency stabilization by means of nonlinear Ramsey resonances,  Appl. Phys. B \textbf{48}, 165-171 (1989).
\bibitem{Letchumanan:2006} V. Letchumanan, P. Gill, A.G. Sinclair and E. Riis, Optical-clock local-oscillator stabilization scheme, J. Opt. Soc. Am. B \textbf{23}, 714 (2006).
\bibitem{Yudin:2016} V.I. Yudin, A.V. Taichenachev, M.Yu. Basaleev, Synthetic frequency protocol in the Ramsey spectroscopy of clock transitions, arXiv:1602.00331 (2016).

\end{thebibliography}
\end{document}